\documentclass[%
 preprint,
prb,
]{revtex4-1}

\usepackage{amsmath}  
\usepackage{amsfonts} 
\usepackage{graphicx}
\usepackage{dcolumn}
\usepackage{bm}
\usepackage{xcolor}
\usepackage{siunitx}
\usepackage{braket}
\usepackage{longtable}
\usepackage{booktabs}
\usepackage{url}
\usepackage{hyperref}
\hypersetup{
colorlinks=true, 
urlcolor=blue,
linkcolor=black}
\usepackage{siunitx}
\ifdefined\qty\else
  \ifdefined\NewCommandCopy
    \NewCommandCopy\qty\SI
  \else
    \NewDocumentCommand\qty{O{}mm}{\SI[#1]{#2}{#3}}
  \fi
\fi
\ifdefined\unit\else
  \ifdefined\NewCommandCopy
    \NewCommandCopy\unit\si
  \else
    \NewDocumentCommand\unit{O{}m}{\si[#1]{#2}}
  \fi
\fi

\begin{document}


\title{An instructional lab apparatus for quantum experiments with single nitrogen-vacancy centers in diamond
}


\author{Zhiyang Yuan}%
 \affiliation{Department of Electrical and Computer Engineering, Princeton University, Princeton, New Jersey 08544, USA}
\author{Sounak Mukherjee}%
 \affiliation{Department of Electrical and Computer Engineering, Princeton University, Princeton, New Jersey 08544, USA}
\author{Aedan Gardill}%
 \affiliation{Department of Physics, University of Wisconsin, Madison, Wisconsin 53706, USA}
\author{Jeff D. Thompson}%
 \affiliation{Department of Electrical and Computer Engineering, Princeton University, Princeton, New Jersey 08544, USA}
\author{Shimon Kolkowitz}%
 \affiliation{Department of Physics, University of Wisconsin, Madison, Wisconsin 53706, USA}
 \affiliation{Department of Physics, University of California, Berkeley, California 94720, USA}
\author{Nathalie P. de Leon}%
 \email{npdeleon@princeton.edu}
 \affiliation{Department of Electrical and Computer Engineering, Princeton University, Princeton, New Jersey 08544, USA}

\date{\today}


\begin{abstract}
Hands-on experimental experience with quantum systems in the undergraduate physics curriculum provides students with a deeper understanding of quantum physics and equips them for the fast-growing quantum science industry.
Here we present an experimental apparatus for performing quantum experiments with single nitrogen-vacancy (NV) centers in diamond. This apparatus is capable of basic experiments such as single-qubit initialization, rotation, and measurement, as well as more advanced experiments investigating electron-nuclear spin interactions. We describe the basic physics of the NV center and give examples of potential experiments that can be performed with this apparatus. We also discuss the options and inherent trade-offs associated with the choice of diamond samples and hardware. The apparatus described here enables students to write their own experimental control and data analysis software from scratch all within a single semester of a typical lab course, as well as to inspect the optical components and inner workings of the apparatus. We hope that this work can serve as a standalone resource for any institution that would like to integrate a quantum instructional lab into its undergraduate physics and engineering curriculum.   

\end{abstract}

\maketitle


\section{Introduction}

The growing field of quantum science and engineering has created a need for an up-to-date curriculum, including hands-on quantum lab experiences in controlling and measuring quantum bits (qubits). 
Nitrogen-vacancy (NV) centers in diamond are a promising platform for undergraduate quantum labs \cite{Zhang2018EnsembleNVEsr, sewani2020EnsembleNVEdu} because they are robust and stable qubits at room temperature, the experimental setup does not require vacuum or cryogenic systems like many other atomic or solid-state qubits, and they can be used to study a variety of quantum phenomena. 

Here we outline a design for an instructional lab apparatus that enables students to perform single-qubit initialization, rotations, and readout, as well as exploring electron-nuclear spin interactions that can be used to teach about two-qubit gate operations. Compared to a typical experimental setup,\cite{Maze2008} this apparatus has a simplified and robust optical design, combined with a scanning sample stage to allow for the optical alignment to be static, a novel microwave printed circuit board design enabling high Rabi frequency while avoiding sample wire bonding and remounting, and a pulse synthesis and measurement chain that can be easily programmed. Using this apparatus, we demonstrate optically detected magnetic resonance (ODMR), coherent control of the NV center ground states, and study interactions between the NV center electronic spin and a nearest-neighbor $^{13}$C nuclear spin. We also describe a few extensions of these experiments.

Using the apparatus described below, students can start from a prepared setup and a blank control software notebook (see an example in the Supplementary Materials, Appendix \ref{appendix:example_code}), write and test their own control programs, perform experiments on the setup, and analyze and report their data. In our experience, all of these steps can be completed by a dedicated student within a typical semester.
With different options, the total cost of the setup can vary from \$68,000 to \$100,000.

\section{NV centers in diamond}

\begin{figure*}[h!]
    \centering
    \includegraphics{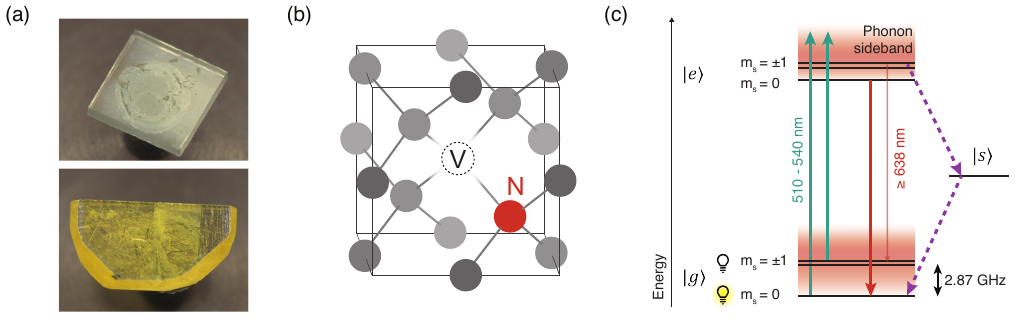}
    \caption{(Color online) NV centers in diamond. (a) Images of diamond samples used for quantum experiments. Top: electronic grade single-crystal chemical vapor deposition (CVD) diamond, size: $3.0 \times 3.0 \times 0.5$ mm. Bottom: High-pressure high-temperature (HPHT) diamond, size: $6.3 \times 3.4 \times 0.5$ mm. The yellow color of the sample arises from nitrogen impurities. (b) Ball and stick model of the NV center atomic structure, showing the vacancy (dashed circle), substitutional nitrogen (red circle), and carbon atoms (gray circles). (c) Simplified energy level diagram of the negatively charged NV center showing the spin triplet ground ($\ket{g}$) and excited ($\ket{e}$) states and metastable singlet state ($\ket{s}$).}
    \label{fig:nv_physics}
\end{figure*}

NV centers are point defects in diamond formed by replacing two carbon atoms with a substitutional nitrogen next to a lattice vacancy [Fig.~\ref{fig:nv_physics}(b)].\cite{doherty2013} The electronic states associated with an NV center are tightly localized to atomic length scales; i.e. they are well-described by linear combinations of molecular orbitals.\cite{larsson2008electronic, gali2008ab} The NV center can have different charge states,\cite{Aslam2013} and the neutral NV center often captures an electron from surrounding defects, resulting in an electronic configuration with a total of six electrons. This charge configuration, the negatively charged NV center, is widely used as a qubit because of its useful spin and optical properties. Unless otherwise noted, the term NV center here refers to the negative charge state.

The two free electrons in the highest energy single-particle orbitals of the NV center form a spin-1 system, with a simplified energy level diagram shown in Fig.~\ref{fig:nv_physics}(c). Transitions between the spin triplet ground states ($\ket{g}$) and the excited states ($\ket{e}$) are enabled by spin-preserving dipole interactions. The ground and excited states are separated by 1.945~eV (637~nm),\cite{zaitsev2013diamondopticalbook, davies1976optical} and electron-phonon coupling gives rise to a broad phonon sideband that spans over 150 nm in both absorption and emission. Most room temperature experiments with NV centers use wavelengths between 510 nm and 540 nm for excitation,\cite{Aslam2013} and collect fluorescence between 637 nm and 800 nm.\cite{alkauskas2014fluorLineshape}

Importantly, this level structure of the NV center allows for optical readout and initialization of the spin state at room temperature.  In the excited state, the $m_s=\pm1$ spin states preferentially undergo an intersystem crossing to the singlet state ($\ket{s}$),\cite{tetienne2012} which has a lifetime around 12 times longer than the excited triplet state.\cite{manson2006nvElectronicStructure, robledo2011spin} For a readout window of 300 ns (related to the singlet state lifetime: 178 ns at 300 K \cite{robledo2011spin}), an NV center initially in the $m_s = 0$ state will cycle through the excited state multiple times and emit photons, while an NV center initially in $m_s = \pm1$ states will emit 30\% fewer photons on average because of the preferential intersystem crossing to the shelving singlet state. This allows for optical readout of the NV center spin state by collecting the spin-dependent fluorescence.\cite{manson2006nvElectronicStructure} On the other hand, a sufficiently long green excitation pulse ($>$ \qty{1}{\micro\second}) will cycle the NV center through the intersystem crossing multiple times and preferentially drive the spin to the $m_s = 0$ state, achieving a steady state population of 80\% in the $m_s=0$ state.\cite{harrison2004opticalpolarization, robledo2011spin} This forms the basis of NV center spin initialization.

In the spin triplet ground states, the $m_s=0$ state is separated from the $m_s=\pm1$ states by a zero-field splitting of 2.87 GHz due to the spin-spin interaction.\cite{lenef1996electronic} The Zeeman effect splits the $m_s=\pm1$ states under an external bias magnetic field. The $m_s=0$ state and one of the $m_s=\pm1$ states are typically chosen as the qubit states, which can be manipulated by microwave pulses.

Because of the high Debye temperature and low noise environment of the diamond host material, NV centers can have long spin relaxation times ($T_1$ up to 6
 ms) \cite{cambria2023} and spin coherence times ($T_2$ up to 3 ms) \cite{bar-gill_solid-state_2013} at room temperature.

The electron spin can also interact with nearby nuclear spins, enabling multi-qubit operations and creating entanglement between qubits. For example, the nuclear spin of the nitrogen atom in the NV center causes hyperfine splittings of the NV center spin states. On the other hand, the natural 1.1\% abundance of $^{13}$C in diamond allows us to find NV centers with strong coupling to nearby nuclear spins. Coherent interactions between the NV center electronic spin qubit and neighboring nuclear spin qubits have been demonstrated.\cite{dutt2007quantum, neumann2010single} Here we show that these properties of the NV center can be explored in an undergraduate teaching lab setting, with an appropriately designed experimental apparatus.

\begin{figure*}[h!]
    \centering
    \includegraphics{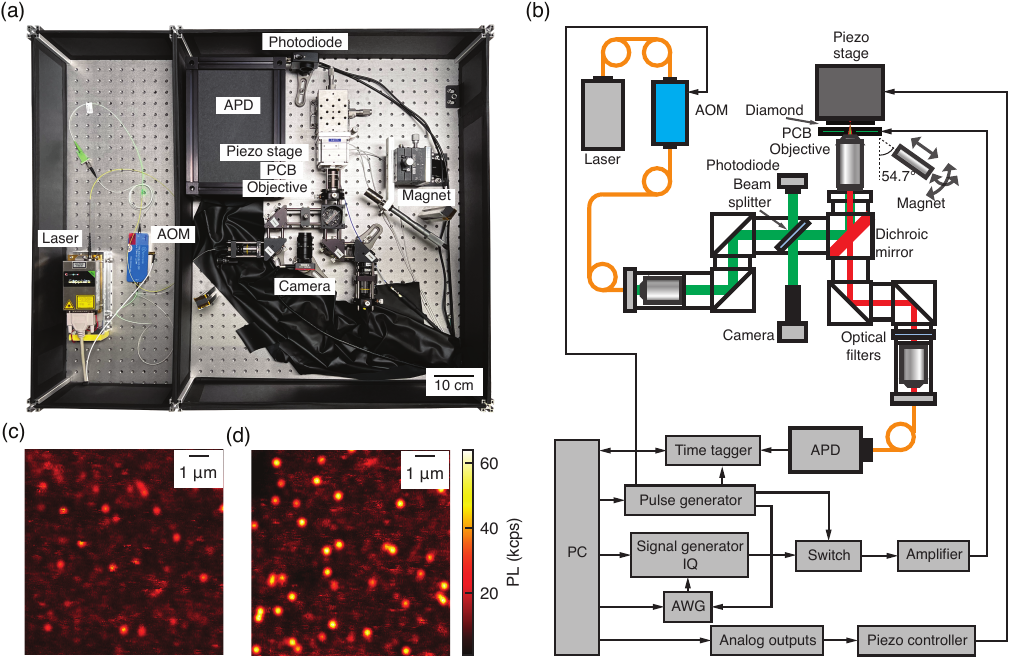}
    \caption{(Color online) Educational quantum laboratory based on NV centers in diamond. (a) Image of the optical setup of the educational lab. The whole optical setup is built on a $30''\times36''$ optical table. The main optics of the confocal setup are mounted in a cage system. An objective is mounted horizontally for the imaging. The sample is mounted on a piezo stage,s which is further mounted on a manual XYZ translational stage. The permanent magnet is mounted on a pair of goniometers for aligning the bias magnetic field to the NV axis. The whole setup is enclosed by black hardboards and the collection fiber and the APD are covered by an extra layer to minimize the dark counts. (b) Schematic of the NV center educational lab. The arrows represent the directions of data and control signal flows. Electronic devices not shown in (a) are placed on a lab bench next to the optical table. (c-d) Example confocal scans of a commercially available HPHT diamond (c) and an implanted ELSC diamond (d). Bright spots are single NV centers. The power of the \qty{532}{\nano\meter} laser for the scans is \qty{270}{\micro\watt} measured before the objective. The scan range is \qty{10}{\micro\meter}$\times$\qty{10}{\micro\meter} with a step size of \qty{0.05}{\micro\meter}. }
    \label{fig:confocal}
\end{figure*}

\section{System design}
\subsection{Optical setup}

The scanning confocal microscope requires nanometer-scale positioning, a scan range of at least tens of microns to locate resolvable NV centers, and high collection efficiency of fluorescence photons. Figure~\ref{fig:confocal} shows an example from our educational lab with the schematic for the optical and electronic components, which are discussed in detail below.  In this design, we employ a cage mount system, making the optical alignment more stable compared to free-standing optics. We also chose a fiber-coupled acousto-optic modulator (AOM) for modulating the excitation laser, as well as a fiber-coupled avalanche photodiode (APD) for ease of alignment and operation. Another key design choice is in mounting the sample and objective sideways, which allows the beam path to be oriented entirely along the plane of the table, simplifying the optical setup. 

\subsubsection{Confocal microscope design}

We now discuss the components employed in our confocal microscope. First, a fiber-coupled green laser is sent to an AOM to enable fast pulsing of the laser, with $\sim20$~ns rise and fall times. The output of the AOM is then sent into the cage system through a collimated objective lens.

A 50:50 beamsplitter cube is included in the green path to allow monitoring and alignment capabilities with the green laser. First, the beamsplitter allows the green laser power to be monitored with a photodiode, which, for example, allows the rise and fall time of the laser to be measured and helps visualize the pulse sequences on an oscilloscope. Second, a camera imaging the diamond surface helps with positioning the diamond sample in the focus of the sample objective.

After the beamsplitter, the green path is reflected by a longpass dichroic mirror and focused through a high numerical aperture (NA=$0.9$) air objective onto the diamond sample. Mounted on the objective is a PCB-printed microwave antenna (see Section~\ref{sec:PCB} for more details). The diamond itself is mounted on a 3-axis piezo stage to position the diamond with respect to the objective's focus. The piezo stage is further mounted on a manual translation stage to enable a larger translation range during sample mounting. 

Additionally, a permanent magnet is placed near the diamond sample, which breaks the degeneracy between the $m_s = +1$ and $-1$ spin ground states. The magnet is mounted on a micrometer head for adjusting the distance to the sample to control the magnetic field strength at the NV center. The micrometer is held by a \qty{54.7}{\degree} magic angle adapter to roughly align the magnetic field with the NV axis in a (100) oriented diamond, which is the most commonly available orientation. Together they are mounted on a goniometer pair to control the magnetic field orientation. The center of the rotation is placed at the sample position, allowing independent control of the field strength and orientation. 

The longer wavelength fluorescence (637--800 nm) from the NV centers is collected by the same high NA air objective. The red fluorescence path is separated from the green excitation path with the dichroic mirror, and a 633 nm longpass filter additionally filters out excitation light. A confocal microscope achieves diffraction-limited resolution by using a spatial pinhole to collect light from a focal spot in the sample. In this apparatus, mode filtering is accomplished with a single-mode fiber instead of a pinhole to simplify the optical alignment and achieve better background rejection. Examples of scanning confocal images of individual NV centers in our educational lab are shown in Figs.~\ref{fig:confocal}(c) and (d).

We provide a procedure for aligning the optics of the confocal microscope in 
the Supplementary Materials (Appendix~\ref{appendix:alignment}). Additionally, we provide a full list of suggested parts to build a confocal microscope and the associated hardware for making NV center measurements in
the Supplementary Materials (Appendix~\ref{appendix:parts}).

\subsection{PCB design for microwave delivery\label{sec:PCB}}
\begin{figure*}[h!]
    \centering
    \includegraphics{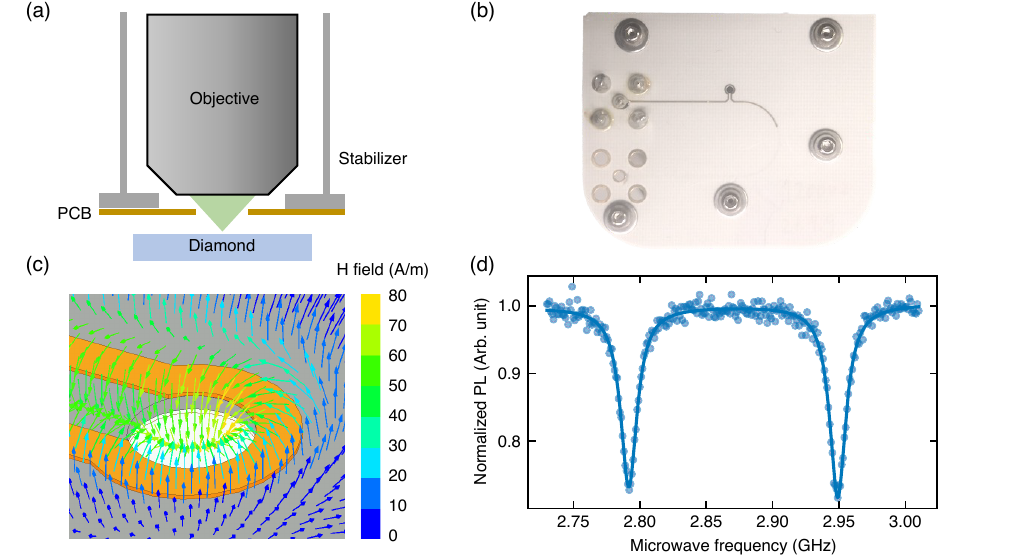}
    \caption{(Color online) PCB design for manipulating NV center spin states. (a) Schematic of the PCB configuration around the air objective. The central opening hole of the PCB allows the green excitation laser and NV center fluorescence to pass through. The stabilizer for the PCB has enough clearance for the objective and the length of it is machined precisely to control the distance between and PCB and the diamond. (b) Picture of a PCB used in the experiment. The gold copper trace forms a loop around the central opening hole of the PCB. One end is soldered to a SMP connector and the other end is left open in this configuration. Five screws and washers are used to mount the PCB to the stabilizer. (c) HFSS simulation of the magnetic field generated by the PCB. The vector magnetic field on a sheet \qty{200}{\micro\meter} above the copper trace is plotted and the field is perpendicular to the PCB at the center of the opening hole. The simulation is done at \qty{2.8}{\giga\hertz} and with an excitation power of \qty{1}{\watt}. (d) Example of the ODMR data for a single NV center at around 28 G. This data shows the transitions between $m_s=0$ and $m_s=\pm1$ states driven by the magnetic field induced by the PCB. The left and right dips are at \qty{2.7917}{\giga\hertz} and \qty{2.9490}{\giga\hertz} respectively. The blue points are experimental data and the blue line is the fitted curve to a two Lorentzian function.}
    \label{fig:PCB}
\end{figure*}

To coherently manipulate the NV center spin state, microwave pulses are generated and delivered to the NV center location. The splitting between spin $m_s=0$ and $m_s=\pm1$ states is \qty{2.87}{\giga\hertz} at zero magnetic field, and the transition frequencies can be shifted by the Zeeman effect. Microwave tones at the transition frequencies can be created by a signal generator and further amplified by a microwave amplifier. Additionally, timed pulsing of the signal generator output is achieved with a microwave switch, allowing pulses as short as \qty{10}{\nano\second}. To drive the NV center spin ground state transitions, the AC magnetic field should have a component perpendicular to the NV axis and the strength should be high enough so that the driving is faster than the spin dephasing rate (usually in the microsecond range). For example, a thin copper wire or loop is commonly used \cite{childress2006coherent, bucher2019quantum} and others have demonstrated microwave delivery using lithographically fabricated strip lines.\cite{sangtawesin2019origins} The PCB design described here offers several key advantages, including compatibility with an air objective mount, unobstructed laser transmission, high microwave power delivery, and long-lasting durability.

By mounting the PCB directly on the air objective, we simplify the experimental setup and ensure consistent microwave delivery at the center of the optical field of view, regardless of the diamond sample positioning. Here, we design the PCB with a thru-hole such that the excitation and imaging can always pass through unobstructed during a confocal scan, enabling efficient excitation and fluorescence detection. We select the thinnest available PCB substrate (Rogers RO4350B, $0.004''$ thickness) and mount the PCB in between the objective and the sample. The mount is designed to stabilize the thin PCB and keep it 100 to \qty{200}{\micro\meter} away from the sample. The diameter $d$ of the opening hole is set to \qty{1}{\milli\meter}, which should be greater than the beam diameter at the PCB position:
\begin{equation}
    d > 2 l\times NA
\end{equation}
where $l$ is the distance between the PCB and the sample.

In addition to optical imaging compatibility, the PCB design is optimized to provide sufficient microwave driving power. The layout and construction of the PCB incorporate efficient microstrip transmission lines, ensuring low insertion loss and impedance matching for effective microwave power delivery. This feature enables less than \qty{100}{\nano\second} $\pi$-pulse times with the microwave source power set to 0 dBm (amplifier gain: 45 dB, see 
Supplementary Materials, Table~\ref{tab:microwave-equipment}), which is sufficient for most experiments. To further increase the driving AC magnetic field strength, we can leave one port of the PCB open to form standing waves along the microstrip transmission line. By varying the length of the trace, antinodes can form around the center loop, which can in principle increase the magnetic field amplitude by a factor of two. In Fig.~\ref{fig:PCB}(b), we show an example of a PCB with one end of the trace open and the length of the trace optimized for the magnetic field generation around the loop. With this PCB, less than \qty{50}{\nano\second} $\pi$-pulse time was achieved when the microwave source power was set to 0 dBm.

Furthermore, our PCB design is very robust against sample impingement and can act as a protection for the objective against sample crashes. By mounting the PCB on the objective, we can leave the PCB untouched when switching samples. We observed no degradation in microwave power delivery over the course of 1.5 years. The design files for the PCB and mount can be found in 
the Supplementary Materials (Appendix \ref{appendix:example_code}).

\subsection{Pulse generation and measurement}

\begin{figure}[h!]
    \centering
    \includegraphics{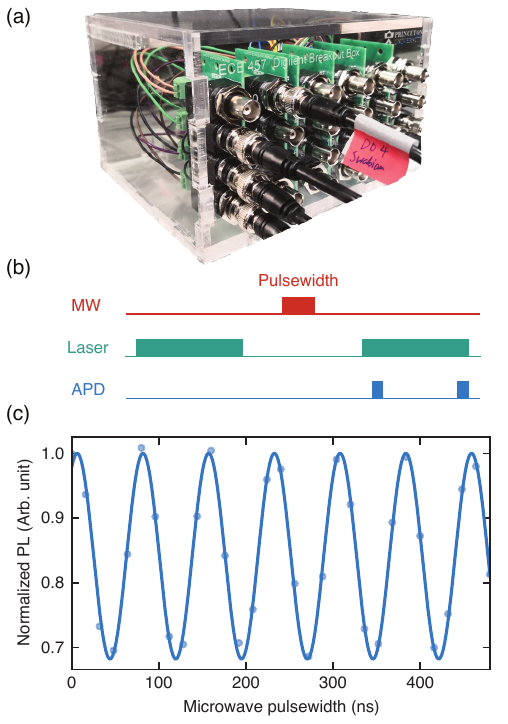}
    \caption{(Color online) a) Breakout box for the Digilent Analog Discovery 2 card. Channels of the card are connected to 24 BNC connectors for ease of connecting with other devices. (b) Pulse program of the Rabi oscillation experiment. During the experiment, the pulsewidth of the microwave channel will be varied. The whole sequence will be repeated multiple times. The green readout pulse of the previous sequence can be used as the initialization pulse for the next sequence to shorten the total experimental time. The delay between the signal counter channel and the green channel is from the AOM delay time. (c) Example of Rabi oscillation data. The blue points are experimental data and the blue line is the fitted curve to a sinusoidal function: $PL(t) = A \sin(\omega t+\phi) + B$, where $\omega$ is the Rabi frequency. A Rabi $\pi$-pulse time of \qty{38}{\nano\second} is demonstrated with 16 W microwave power.}
    \label{fig:Pulse generator}
\end{figure}

In lab experiments, we design synchronized pulse sequences to trigger the excitation laser AOM, the microwave switch, and counter channels. For this task, we employ the Digilent Analog Discovery 2 card, a multi-function instrument with 16 channels for digital pattern generation (3.3 V CMOS, 100 MS/s), which is sufficient for the timing control of most NV center experiments. Additionally, the two programmable power supply channels (0.5 V to 5 V, -0.5 V to -5 V) can provide the $\pm$\qty{5}{\volt} bias voltages for the microwave switch used in this lab. The Analog Discovery 2 card comes with pins for different channels and we build a custom BNC breakout box for easy connections to other devices (see Fig.~\ref{fig:Pulse generator}). The biggest limitation of this device is that it only has a buffer size of 1024 samples for custom pattern generation. This gives a maximal sequence length of around \qty{10}{\micro\second} when \qty{10}{\nano\second} resolution is used, limiting the ability to measure long timescale dynamics ($T_1$, $T_2$) with high timing resolution. For longer pulsed experiments, we used a PulseBlaster card (see 
Supplementary Materials, Table~\ref{tab:pulse generation}). While the PulseBlaster card enables better timing resolution and longer pulse sequences, it is more difficult to implement in control software. For our apparatus, we perform experiments during the semester with the Digilent Analog Discovery 2 card, and we employ a PulseBlaster with pre-written control software for specific final projects and independent projects.

To record APD photon counts we use the Swabian Time Tagger 20, which has 8 input channels with a dead time of \qty{6}{\nano\second}. The channels can work as either sources or triggers for the counters, which allows us to define the readout window relative to other pulses. Multiple available input channels also make it easy to implement correlation measurements like $g^{(2)}(\tau)$ to confirm single photon emitters. 

We also tested a NI DAQ card for data acquisition when multi-channel measurement was not required. This card has multiple functions such as edge counting and analog/digital input/output and is commonly used for NV center experiments (see 
Supplementary Materials, Table~\ref{tab:counter}). Counter channels can be used to read out the APD count rate. We also use analog output signals to control the positions of the piezo system. These output channels and counting channels can be triggered and synchronized by sharing the same hardware clock, which enables fast confocal scans.

\section{Samples}
The primary sample used in the experiments below is an electronic-grade single-crystal (ELSC) diamond implanted with nitrogen ions and processed to form NV centers. However, other readily available samples can also be used. The electronic-grade diamond used in our experiments was implanted with a $^{15}$N ion dose of \qty{1e9}{\per\centi\meter\squared} at \qty{30}{\kilo\electronvolt} and \qty{0}{\degree} tilt angle by INNOViON Corporation. After ion implantation, the sample was annealed at \qty{800}{\celsius} in a home-built vacuum furnace to form NV centers, and then the surface was annealed under an oxygen environment at \qty{455}{\celsius} to generate an oxygen terminated surface. Ion implantation enables the formation of single NV centers at controlled densities and depths, which makes it easier for students to find NV centers and run experiments. We also achieve long coherence times by eliminating defects at the surface by high quality oxygen termination.\cite{sangtawesin2019origins} 

HPHT diamonds contain individually resolvable native NV centers despite their high nitrogen concentration, and can also be used to probe single centers [Fig.~\ref{fig:nv_physics} (a), bottom]. These are more cost effective and readily available, but have the disadvantage of shorter NV center spin coherence times limited by their high nitrogen background. 

All the experiments outlined here can also be readily performed with native NV centers in high purity, off-the-shelf, electronic-grade CVD diamonds from Element Six [Fig.~\ref{fig:nv_physics} (a), top]. Compared with shallow implanted NV centers, natively occurring NV centers far below the surface typically have longer coherence times (Hahn echo $T_2 >$  \qty{500}{\micro\second}). The main experimental complication in using such samples is that the density of NV centers can be nonuniform and sparse in some regions, possibly with densities as low as one NV center in a \qty{40}{\micro\meter}$\times$\qty{40}{\micro\meter} scan range, requiring extensive mapping to find a single NV center. The NV center density can be readily increased by electron irradiation and thermal annealing.\cite{davies1976optical, schwartz2012effects}

\section{Example labs\label{sec:example_labs}}

\subsection{Confocal microscopy}
As a first step to investigate NV centers in the diamond, students learn to find and map single centers by running a confocal scan of the sample. In this lab, students will be asked to obtain a confocal image of the sample with individually resolvable NV centers and fit their data to get accurate positions of the NV centers. Due to the time limit in a typical one-semester course, the confocal microscope has been set up and aligned by the instructors before the start of the course. If time permits, students can be given exercises such as collimating the laser into and out of a single mode fiber, which will help them learn optical alignment. See 
Supplementary Materials (Appendix \ref{appendix:alignment}) for the alignment procedures of this setup. First, students need to move the sample to the focus of the high NA objective. For measuring shallow implanted NV centers, we can simply focus the laser on the front surface of the diamond, which can be checked by the reflection of the green laser on the camera (Fig. \ref{fig:confocal}). Specifically, one can focus the excitation beam on the sample by achieving the smallest spot size on the camera. This is done by moving the sample with respect to the objective with the help of the manual translation stage on which the piezo stage is mounted. For a sample with deep native NV centers, one can start by focusing the laser on the surface and using the piezo stage to move a few micrometers into the diamond to find NV centers.

Second, students learn to synchronize the driving of the piezo stage with the APD photon counting to acquire their confocal images. Our setup uses analog voltages from DAQ channels to set the piezo stage to a particular $(x, y, z)$ position. Students can sweep the positions and read out photon counts from the APD with the same time intervals to perform a raster scan of the sample. This allows them to resolve the NV centers as diffraction-limited bright spots, as shown in Figs.~\ref{fig:confocal}(c) and (d). To run experiments on a particular NV center, the focus can be further adjusted with the APD counts according to the depth of the defect. Students can also track the centers by performing fine scans in  $x$, $y$, and $z$ directions and fitting the data to Gaussian curves to move to the peak positions. This is crucial to account for thermal drifts of the optics that present as drifts in the position of the sample over time, which must be corrected for further experiments. 

\subsection{Single NV center photon statistics}

\begin{figure}[h!]
    \centering
    \includegraphics{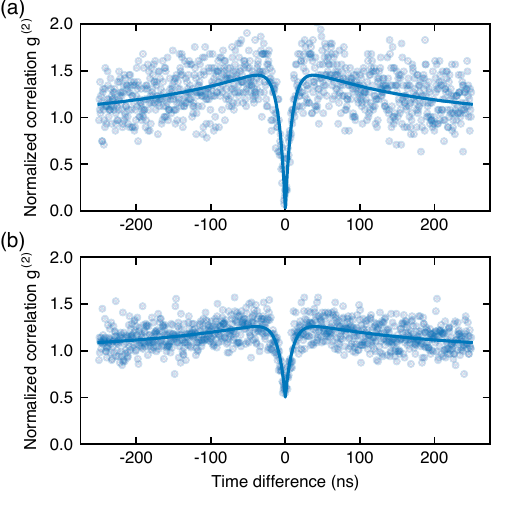}
    \caption{(Color online) The $g^{(2)}$ measurements of (a) a single NV center and (b) multiple NV centers. $g^{(2)}(0)<0.5$ verifies that the fluorescence emission is from a single NV center. The green laser power is \qty{270}{\micro\watt}. The histogram bin width is set to \qty{0.5}{\nano\second} and the measurement is averaged for \qty{5}{\minute}. The blue points are experiment data and the blue lines are fitted curves described in the main text.}
    \label{fig:g2}
\end{figure}

To verify that the bright spots observed in confocal scanning images are single NV centers, we can check the fluorescence photon statistics by measuring the second-order correlation function from each bright spot, which is defined as 
\begin{equation}
    g^{(2)}(\tau) = \frac{\langle I(\tau)I(0)\rangle}{|\langle I\rangle|^2}
\end{equation}
where $I(\tau)$ is the photon intensity at time $\tau$. In this lab, students are asked to modify the setup, obtain the $g^{(2)}$ correlation data of a single NV center, and perform data analysis to study the NV spin dynamics. This can be used as an optional lab as it requires an additional APD and a multichannel time-to-digital converter. 

As a single photon emitter, each single NV center can not emit more than one photon at a time. By creating a histogram of the time differences between two photons, we should observe the probability approaching 0 as the time difference approaches 0. Since $g^{(2)}(0)=1/2$ for two emitters, observing $g^{(2)}(0)<1/2$ is sufficient to prove that the measured spot is a single NV center.  As the dead time of the APD (\qty{22}{\nano\second} for the Excelitas SPCM-AQRH-43-FC) prevents us from measuring these statistics, the fluorescence photons are divided into two APDs by a splitter forming a Hanbury Brown and Twiss (HBT) interferometer (a detailed table of suggested components can be found in 
the Supplementary Materials (Table~\ref{tab:counter} and \ref{tab:APD}). We send the signals from the two APDs to two channels of the Time-to-digital converter (Swabian Time Tagger 20). The Swabian Time Tagger model provides the `Correlation' measurement class for this type of measurement, which makes the implementation more straightforward. An example of a $g^{(2)}$ measurement on a single NV center is shown in Fig.~\ref{fig:g2}. The measured $g^{(2)}$ also exhibits bunching at short times arising from the singlet shelving state. The shelving state traps the NV population and prevents subsequent photon emission up to the shelving state lifetime. These data can be fit to
\begin{equation}
    g^{(2)}(\tau) = \rho^2 [1 - \beta \exp(-\gamma_1 \tau) + (\beta-1) \exp(-\gamma_2\tau)] + (1 - \rho^2)
\end{equation}
where $\rho$ is the ratio between the NV fluorescence and the total signal including the incoherent background, and $\beta$, $\gamma_1$, and $\gamma_2$ come from the transition rates among NV states.\cite{berthel2015photophysics} 
Quantitatively fitting the $g^{(2)}$ data to this model to understand the NV spin and photodynamics, as well as measuring the laser power dependence of these dynamics are accessible extension projects.

\subsection{Optically detected magnetic resonance}
After locating individual NV centers, students can move on to drive the NV spin ground states using microwaves and observe ODMR signals. Students need to familiarize themselves with the microwave setup, write programs to sweep the frequency of microwave signals, understand the optical readout of the NV spin state, and collect data to find the transition frequencies of the NV center. This experiment is done by applying microwaves through the PCB antenna and collecting the NV center fluorescence count rate as a function of the microwave frequency. This can be programmed in continuous (which applies the microwave and laser at the same time) as well as pulsed mode (where the laser and microwave are applied at separate times) using a microwave switch and AOM to pulse the microwaves and laser, respectively. The pulses are generated using the Digilent Analog Discovery 2 card as described before. Continuous green excitation initializes the spin into $m_s=0$, which is the bright state, while a microwave drive redistributes the population with $m_s=\pm1$ state. Because spin $m_s=\pm1$ states have a lower photon count rate, there is a dip in fluorescence when the applied microwaves are on resonance with the spin transition. At zero field, the $m_s=\pm1$ states are degenerate and we observe a single dip at the zero-field splitting (2.87 GHz). At non-zero fields, $m_s=\pm1$ undergo Zeeman splitting, and we observe two dips as shown in Fig. \ref{fig:PCB}(d). Students will perform data fittings to get the transition frequencies. The magnetic field sensed by the NV center can be quantified from the splitting. We can set the microwave frequency to be resonant with one of the $\ket{m_s=0}\leftrightarrow\ket{m_s\pm1}$ transitions to perform single-qubit gate operations as described in the next subsection.

\subsection{Single-qubit experiments}
Using their knowledge of the OMDR spectrum, students can now perform single-qubit operations on the NV center electronic spin and observe spin dynamics by programming pulsed experiments. In this lab, they will be asked to design pulse sequences for Rabi, Ramsey, and Hahn echo experiments, optimize parameters in the pulse sequences, and run data analysis to calibrate single-qubit gates and obtain NV spin characteristic times, $T_2^*$ and $T_2$.  We utilize $m_s=0$ and $m_s=-1$ as the two fiducial qubit states. All pulsed experiments include an initialization into $m_s=0$ state with green excitation and end with a readout using the same. The readout is optimal for a counting window comparable to the singlet state lifetime, and optimizing the counting window is a good exercise to include within this module. Students can calibrate the delay between the pulse applied to the AOM and the actual laser pulse, as well as the delay of the microwave switch, by checking the timing response on an oscilloscope. Then, by varying the microwave pulsewidth and plotting fluorescence photon rate, we can observe Rabi oscillations between the $\ket{0}$ and $\ket{1}$ states of the qubit, as shown in Fig. \ref{fig:Pulse generator}(c). The Rabi frequency is dependent on the microwave power and can be calibrated to obtain the $\pi$-pulse, which is the duration required for complete population transfer from $\ket{0}$ to $\ket{1}$. Sweeping the pulse duration can coherently drive the spin between these two states. Thus we have an effective way to initialize spins and perform single-qubit gate operations.

\begin{figure}[h!]
    \centering
    \includegraphics{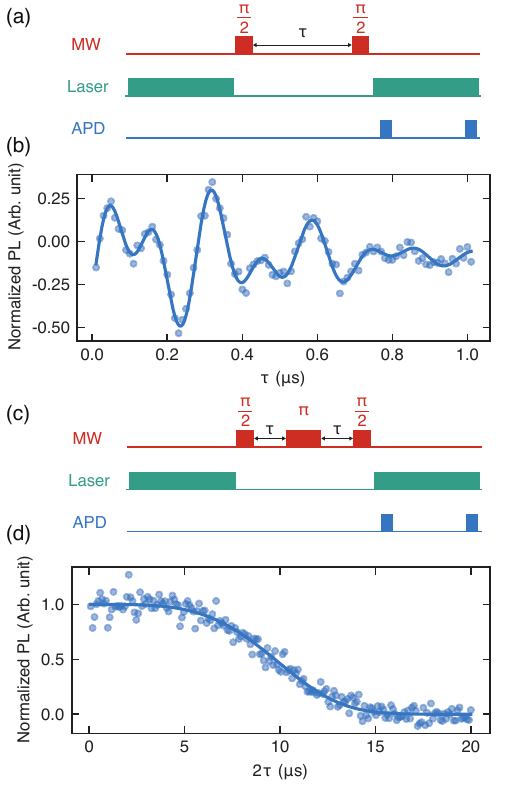}
    \caption{(Color online) (a) Ramsey experiment pulse sequence. (b) NV center Ramsey data. The data (blue points) can be fitted to two frequency tones (blue line): $PL(\tau) = [A_1 \sin(2\pi f_1 \tau + \phi_1)+ A_2 \sin(2\pi f_2 \tau + \phi_2)]\exp[-(\tau/T_2^*)^2] + B$, where $f_1=\qty{7.12}{\mega\hertz}$ and $f_2=\qty{4.22}{\mega\hertz}$ are from the hyperfine coupling with the $^{15}$N nuclear spin.\cite{childress2006coherent} (c) Hahn echo experiment pulse sequence. (d) NV center Hahn echo data. The collapse is due to the interaction with the natural 1.1\% $^{13}$C nuclear spin bath. The blue line is the fitted curve to $\exp[-(2\tau/T_c)^4]$ ($T_c = 10.7\pm0.1$ $\mathrm{\mu s}$).\cite{childress2006coherent} An external magnetic field of 23~G is applied by permanent magnets and is aligned with the NV axis. Revivals of the Hahn echo signal can be observed for longer $\tau$ values due to the rephasing of the $^{13}$C nuclear spin bath. For this, a PulseBlaster is required to program longer pulse sequences.}
    \label{fig:ramsey_hahn}
\end{figure}

Once the $\pi$-pulse is calibrated, further pulsed experiments like the Ramsey and Hahn echo sequences can be programmed to observe free spin dynamics and measure dephasing and decoherence. The Ramsey sequence uses $\pi/2-\tau-\pi/2$ pulses as shown in Fig.~\ref{fig:ramsey_hahn}(a). We start with a superposition of $\ket{0}+\ket{1}$, let it evolve freely for time $\tau$, map the accumulated phase to a population using another $\pi/2$ pulse, and then read out the population using a green pulse. For an ideal two-level system, we expect to see population oscillations with a frequency equal to the microwave detuning. However, the NV center electronic spin is always coupled to the intrinsic nitrogen nuclear spin, so we observe a beating pattern [Fig.~\ref{fig:ramsey_hahn}(b)] from oscillations of multiple two-level systems. In these data, the NV center was created by implanting $^{15}$N ions, thus two frequency tones were observed. For the more common $^{14}$N isotope, one would observe a beating of three frequencies.\cite{childress2006coherent} The decay of the envelope gives us the spin dephasing time, $T_2^*$. This dephasing arises from the slowly varying magnetic fields in the spin bath. 

By introducing a $\pi$ pulse in between, one can implement the Hahn echo sequence [$\pi/2-\tau-\pi-\tau-\pi/2$, Fig. \ref{fig:ramsey_hahn}(c)]. The Hahn echo measurement is commonly used for characterizing qubit coherence times and for NV centers in diamond, this coherence time is usually limited by the $^{13}$C nuclear spin bath. In Fig.~\ref{fig:ramsey_hahn}(d), we plot the first collapse of the NV center spin coherence due to the decorrelation of the $^{13}$C spin bath.  For longer $\tau$, periodic revivals of the NV center spin coherence can be observed. \cite{childress2006coherent} The decay time of the envelope of revivals is defined as the decoherence time $T_2$. More advanced dynamical decoupling sequences like CPMG,\cite{naydenov2011dynamical} XY4,\cite{de2010universal} and so on can also be used to extend this coherence time and they can be programmed by adding an arbitrary waveform generator (AWG) to the setup (Supplementary Materials, Table~\ref{tab:microwave-equipment}).

\subsection{Electron-nuclear spin interactions}
\begin{figure}[h!]
    \centering
    \includegraphics{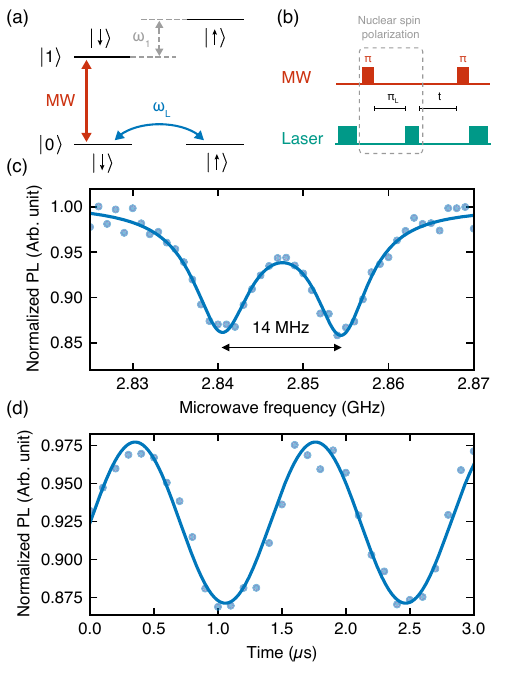}
    \caption{(Color online) (a) Level structure of the NV center electron spin, $\{\ket{0}, \ket{1}\}$, coupled with a neighboring $^{13}$C  nuclear spin, $\{\ket{\downarrow}, \ket{\uparrow}\}$. Microwaves (MW) can drive the population between the electron spin manifold where $\ket{1}$ is split by $\omega_1$ from the hyperfine interaction. The nuclear spin can precess at $\omega_L$ between $\ket{0}\ket{\downarrow}\leftrightarrow\ket{0}\ket{\uparrow}$. (b) Pulse sequence to initialize and read out the nuclear spin by mapping it to the electron spin. (c) ODMR data exhibiting the hyperfine structure associated with nuclear spin, with a splitting of $\omega_1=14$~MHz. The blue points are experimental data and the blue line is the fitted curve to a two Lorentzian function. (d) Nuclear free precession measured via the electron spin using the pulse sequence in (b). The blue points are experimental data and the blue line is the fitted curve to a sinusoidal function.}
    \label{fig:nucelar_prec}
\end{figure}
NV centers in diamond offer a natural platform for studying electron-nuclear spin interactions, which students can access in the same educational lab setup. In this final lab, students will find an NV-nuclear spin pair, study the hyperfine splitting of the NV transition, and design an experiment to see the free precession of the nuclear spin. In typical diamond samples, $1.1\%$ of all carbon atoms are the isotope $^{13}$C, which has a nuclear spin of $I=1/2$. Nearest neighbor $^{13}$C nuclei can have strong enough coupling to allow for well-resolved hyperfine structure \cite{smeltzer201113c} that can be observed in pulsed ODMR measurements [Fig.~\ref{fig:nucelar_prec}(c)] with low enough microwave powers to avoid power broadening of the transition linewidths. Given there are nine lattice sites for $^{13}$C that can give rise to a hyperfine splitting of around \qty{14}{\mega\hertz}, the probability of finding one NV-$^{13}$C pair with this hyperfine splitting is 9\%. Collecting pulsed ODMR spectra of an NV batch should enable finding the desired NV-$^{13}$C pair and pairs with other hyperfine splitting values are in principle also suitable for further experiments. 

The hyperfine splitting of 14~MHz corresponds to the difference in energies of the NV spin $\ket{1}$ state when the $^{13}$C nuclear spin is in  $\ket{\downarrow}$ or $\ket{\uparrow}$ state, as indicated by $\omega_1$ in Fig.~\ref{fig:nucelar_prec}(a). This allows us to selectively drive the NV center electronic spin based on the nuclear spin state. For example, a $\pi$-pulse using a weak microwave drive flips the electronic spin only for the $\ket{\downarrow}$ state due to the large hyperfine splitting. This essentially forms a controlled-NOT or $\mathrm{C_{n}NOT_{e}}$ gate on the electronic spin controlled by the nuclear spin. 

This also allows us to manipulate the nuclear spin by its interaction with the electronic spin.\cite{dutt2007quantum}  To initialize the nuclear spin, we first initialize the NV center electronic spin state in $\ket{0}$ with a green pulse, which leaves the nuclear spin in either of its spin states, $\ket{\downarrow}$ or $\ket{\uparrow}$. This sets the system in a mixed state of $\ket{0}\ket{\downarrow}$ and $\ket{0}\ket{\uparrow}$. A weak MW $\pi$-pulse can drive the $\ket{0}\ket{\downarrow}$ population to $\ket{1}\ket{\downarrow}$, leaving the $\ket{0}\ket{\uparrow}$ population untouched. If an external magnetic field perpendicular to the NV axis is applied, we can observe free precession between the $\ket{0}\ket{\downarrow}$ and $\ket{0}\ket{\uparrow}$ states with Larmor frequency $\omega_L$, while the large hyperfine splitting $\omega_1$ prevents the nuclear precession when the electronic spin is in $\ket{1}$. We wait for a duration $\pi/\omega_L$ or $\pi_L$ for the nuclear spin to flip, followed by another green pulse to get all the population in $\ket{0}\ket{\downarrow}$. This completes the nuclear spin polarization step [Fig.~\ref{fig:nucelar_prec}(b)]. Next, by sweeping the duration between spin polarization and readout pulses, we can observe nuclear free precession as shown in Fig.~\ref{fig:nucelar_prec}(d). The readout of the nuclear spin population can be done in a similar way using a MW $\pi$-pulse and green pulse to map it back to the electron spin.\cite{dutt2007quantum} 

In this experiment, students study the electron-nuclear spin interaction by driving and measuring the NV electron spin. To extend this to a two-qubit gate, the setup would require the addition of a separate RF tone to drive the nuclear spin directly, which has been demonstrated in the literature,\cite{jelezko2004observation} but entails several technical complications associated with the high RF power required such as thermal drift.

\section{Conclusion}
In this paper, we have described the design and implementation of a confocal microscope for implementing quantum experiments of single NV centers in diamond including single-qubit gates and electron-nuclear spin interactions. We have chosen hardware such that in a single academic term, students should be able to program control software from scratch, helping to give them hands-on experience in quantum measurement and control. The apparatus also allows for a multitude of potential independent projects on quantum sensing, quantum control, and quantum information processing with single NV centers and small registers of nuclear spin qubits. We hope that this serves as a resource for any institution seeking to develop innovative laboratory courses into their undergraduate physics and engineering curriculum.  

\begin{acknowledgements}
This work was primarily supported by the National Science Foundation (QuSEC-TAQS OSI 2326767 and QLCI Grant No. OMA-2120757), and SK was supported by National Science Foundation Grant No. ~2143870.
\end{acknowledgements}

\section*{Author declarations}
\subsection*{Conflict of Interest}
The authors have no conflicts to disclose.

\bibliography{apssamp}

\newpage
\section{Supplementary material}
\appendix

\section{Confocal optics alignment\label{appendix:alignment}}
Here we provide instructions to align the optics of the confocal microscope. It is assumed that the cage system is assembled and that there is a sample with NV centers to use for imaging. 

The alignment procedure follows five main steps:
\begin{enumerate}
    \item Collimating the green excitation path
    \item Collimating the red collection path
    \item Centering the beam through the sample objective
    \item Overlapping the excitation and collection paths
    \item Making fine adjustments using an NV center
\end{enumerate}

Before starting, if there is currently a diamond sample in the stage, retract it from the objective to remove the sample. Remove the sample objective from the cage system as well. Place a drop-in mirror to the cage system somewhere between the dichroic mirror and sample objective to send the beam path out of the cage system.

\subsubsection{Collimating the green excitation path}
In this step, you will collimate the excitation beam, meaning that the beam's diameter will be roughly the same throughout the confocal microscope optical path without any significant divergence or convergence. A straightforward approach for roughly collimating the beam is to ensure the beam remains the same size (by eye) across a long distance.

\begin{enumerate}
    \item Using a drop-in mirror placed in the cage system, direct the beam path out of the cage system and send it across the room. Make sure the green laser is set to a safe, low power and notify all other people in the same room.
    \item Monitor the beam's diameter as it exits the cage system and far away from the cage system. Adjust the relative position of the fiber to the coupling optic of the excitation path so the size of the beam is the same at these two points.
    \item Make sure the beam does not come to a focus between these two points, the beam's diameter should remain the same along its beam path.
\end{enumerate}

\subsubsection{Collimating the red collection path}

Next, the same process is performed for the collection path. A red laser is sent backwards through the fiber that connects to the APD. It is helpful to turn off the green laser during this step. Importantly, make sure the APD is turned off to protect it from damage.

\begin{enumerate}
    \item Unplug the fiber that leads to the APD and plug it into a red laser source. Turn on the red laser source.
    \item Depending on the long pass filter used in the collection path, the red laser source may be blocked by the filter. If that is the case, remove the filter for the alignment.
    \item Following similar steps as with the green light, adjust the relative position of the coupling lens and the fiber so that the red beam is uniform in size along its path over a large distance. 
\end{enumerate}

\subsubsection{Centering the beam through the sample objective}

In this step, we align the excitation path to pass through the center of the objective perpendicularly. For this purpose, we put two irises on the ends of a slotted lens tube in place of the objective to allow alignment.  

\begin{enumerate}
\item Remove the drop-in mirror so the beam path continues through the cage system.
    \item Attach irises on the ends of a slotted tube. A longer tube will allow more precise alignment.
    \item Screw this assembly in place of the objective.
    \item Adjust the mirrors in the excitation path to center the beam on the two irises. To center the beam on the first iris (closest to the cage system), adjust the excitation mirror farther away from the irises. To center the beam on the second iris, adjust the closer excitation mirror, which is the dichroic mirror in this lab. Repeat these two steps multiple times until the beam passes through the centers of both irises.
\end{enumerate}

\subsubsection{Overlapping the excitation and collection paths}

The next step is to overlap the excitation and collection paths after the dichroic mirror. It is helpful to put the drop-in mirror back into the cage system and steer the beam across the room. You will ensure the beams are overlapped along their entire path by overlapping the beams at two distinct points that are far apart (for example, right outside the confocal microscope and across the room). 

Since the excitation path has already been aligned to the objective in the previous step, we align the collection path to the excitation path here.

\begin{enumerate}
    \item Turn on both the red handheld laser and the green laser, and roughly match their power. 
    \item Adjust the mirrors in the collection path. Monitor the overlap of the beams near the confocal microscope and far from the confocal microscope. To adjust the collection beam's position near the confocal microscope, adjust the mirror farther away from the dichroic mirror. To adjust the position far from the confocal microscope, adjust the other collection mirror closer to the dichroic mirror. 
\end{enumerate}

\subsubsection{Making fine adjustments using an NV center}

At this point, the confocal microscope should be well collimated and the beam paths should be overlapped. The final step is to use the signal from an NV center to make fine adjustments to the beam paths to ensure the confocal microscope is as well aligned as possible. 

First, make sure the following optics are replaced that were removed for alignment. 

\begin{itemize}
    \item Remove the slotted tube with irises and replace the sample objective. 
    \item If any filters were removed in the collection path, put them back now.
    \item Reconnect the collection fiber to the APD.
    \item Place the diamond sample on the sample stage.
\end{itemize}

The next steps require finding an NV center to use as the source of our signal to make fine adjustments to the excitation and collection paths.

\begin{enumerate}
    \item First, manually move the sample into focus of the sample objective. Monitor the back-reflected image of the sample on the camera, captured by the beam-splitting cube in the excitation path. The back-reflected image of the sample will come to a focused point when the sample objective's focus is focused on the diamond surface.  
    \item Perform confocal scans to find NV centers with the hardware. If you are not able to find NV centers, you may need to repeat the previous steps and ensure your beams are overlapped.   
    \item Once you have identified an NV center, you will need to monitor the fluorescence counts from the NV center under constant excitation.
    \item While monitoring the counts from the NV center, adjust the collection and excitation mirrors and the Z-position of the fibers to maximize the signal. ``Walking" the mirrors is a useful process to make adjustments to a beam's path, in which the horizontal (vertical) mirror knob of one mirror is detuned, and then the second mirror's horizontal (vertical) knob is used to increase the signal.
\end{enumerate}

Once the signal from the NV center has been maximized, the confocal microscope is ready to use for experiments. 

Over time, optics will drift and realignment of the optics may be necessary. It is often sufficient to perform just a subset of these full steps to realign the optics. For example every few months it may be useful to follow the final few steps to make fine adjustments to the mirrors and fiber z-positions to maximize the signal of an NV center. 

\section{Parts list\label{appendix:parts}}
Here we offer suggested components to build an experimental setup to measure single NV centers. For some components of the experiment, multiple options are presented, between which the reader can weigh the benefits and drawbacks of the options. The prices were checked on May 4, 2024, and may vary among different vendors. To lower the cost, used equipment can also be considered.

\begin{table*}[h!]
	\centering
    \caption{Suggested components for the excitation laser. Two options are provided, which offer different features such as internal/external modulation.}
	\begin{ruledtabular}
	\begin{tabular}{p{1.2in}p{0.8in}p{0.8in}p{2in}p{0.5in}p{0.2in}}
	Item & Vendor & Model & Description & Price (USD) & Qty \\
	\hline
\textbf{Option 1:}\\
    Green laser & Hübner Photonics  & 0520-06-01-0080-100 & 520 nm, 80 mW max output, 150 MHz modulation & 3450 & 1 \\
\textbf{Option 2:} \\
    Green laser & Coherent & Sapphire 532 FP & 532 nm, 300 mW  max output, no modulation & 13579 & 1 \\
    Heat sink & Coherent & 1110061 & Sapphire laser heat sink & 489 & 1 \\
	AOM (and driver) & Brimrose & TEM-200-25-20-532-2FP & Allows laser modulation. 20 ns rise time, 200 MHZ center modulation frequency,  25 MHz modulation bandwidth, fiber-coupled & 4200 & 1 \\
	\end{tabular}
	\end{ruledtabular}
	\label{tab:laser}
\end{table*}
\clearpage

\begin{table*}[h!]
	\centering
    \caption{Suggested components for pulse generation to control the timing of experimental hardware. Three options are provided, which offer different sampling rates and number of outputs, and span a wide range of prices. }
	\begin{ruledtabular}
	\begin{tabular}{p{1.2in}p{0.8in}p{0.8in}p{2in}p{0.5in}p{0.2in}}
	Item & Vendor & Model & Description & Price (USD) & Qty \\
	\hline
\textbf{Option 1:}\\
	Pattern generator & Digilent & Analog Discovery 2 & 16 digital output channels: 100 MHz sample rate 
 & 299 & 1 \\
\textbf{Option 2:} \\
        Pulse Streamer & Swabian & PS 8/2 & 8 digital output channels: 1 GHz sample rate.  2 analog output channels: 125 MHz sample rate & 4471 & 1 \\
\textbf{Option 3:}\\
	PulseBlaster & SpinCore & PBESR-PRO-500-PCI & 21 digital output channels: 500 MHz sample rate & 4485 & 1 \\
	\end{tabular}
	\end{ruledtabular}
	\label{tab:pulse generation}
\end{table*}

\begin{table*}[h!]
	\centering
    \caption{Suggested components for data acquisition. Two options are provided, the first offering time tagging capabilities (which allows for the $g^{(2)}$ correlation measurements) and the second acting just as a counter.}
	\begin{ruledtabular}
	\begin{tabular}{p{1.2in}p{0.8in}p{0.8in}p{2in}p{0.5in}p{0.2in}}
	Item & Vendor & Model & Description & Price (USD) & Qty \\
	\hline
\textbf{Option 1:}\\
	Time-to-digital converter & Swabian & Time Tagger 20 & 8.5 M tags/s, 8 input channels, 34 ps RMS jitter & 11300 & 1 \\
\textbf{Option 2:}\\
	Multifunction I/O device & National Instruments & USB-6363 &  4 input counter channels, 32 bit resolution & 4228 & 1 \\
	\end{tabular}
	\end{ruledtabular}
	\label{tab:counter}
\end{table*}
\clearpage

\begin{table*}[h!]
	\centering
    \caption{Suggested components for positioning the sample with respect to the confocal microscope focus. Additionally, micrometer-scale positioning is needed for adjusting the position of the sample.}
	\begin{ruledtabular}
	\begin{tabular}{p{1.2in}p{0.8in}p{0.8in}p{2in}p{0.5in}p{0.2in}}
	Item & Vendor & Model & Description & Price (USD) & Qty \\
	\hline
\multicolumn{5}{l}{\textbf{Nanometer positioning:}} \\
	3-axis piezoelectric stage & Mad City Labs & Nano3D200 \& Nano-Drive  &  XYZ positioning, 1 nm resolution, 200 $\mu$m range of motion, closed-loop & 12222 & 1 \\
 \multicolumn{5}{l}{\textbf{Accessory:}} \\
    USB analog output & MCC & USB-3101 & 4 analog output channels to control XYZ positioning, $\pm$10 V output, 16 bits resolution. Note: if the USB interface option is chosen in the MCL Nano-Drive, instructions through USB can be used to control the positioning instead. & 345 & 1 \\
\multicolumn{5}{l}{\textbf{Micrometer positioning:}} \\
		XYZ Linear Stage & Newport & 562-XYZ & Used in addition to nanopositioning option to manually position diamond under focus. & 2901 & 1 \\
		Micrometer & Newport & SM-13 & Required micrometers for XYZ stage listed above. 13 mm travel & 100 & 3 \\
	\end{tabular}
	\end{ruledtabular}
	\label{tab:positioning}
\end{table*}

\begin{table*}[h!]
	\centering
    \caption{Suggested components for microwave generation, and optional I/Q modulation, which is needed to perform dynamical decoupling sequences.}
	\begin{ruledtabular}
	\begin{tabular}{p{1.2in}p{0.8in}p{0.8in}p{2in}p{0.5in}p{0.2in}}
	Item & Vendor & Model & Description & Price (USD) & Qty \\
	\hline
   Signal generator & SRS & SG384 \emph{( w/ option 3: External I/Q Mod.)} & DC to 4.05 GHz frequency output, -110 to 16.5 dBm power output, I/Q modulation & 7850 & 1 \\
		Arbitrary Wave Generator & Keysight & 33622A & Frequency: 120 MHz, used for I/Q modulation & 8930 & 1 \\
 \multicolumn{5}{l}{\textbf{Other necessary components:}} \\
		Microwave Amplifier & Mini-circuits & ZHL-16W-43-S+  & 1.8 to 4 GHz frequency range, 16 W amplification & 2421 & 1 \\
		Microwave switch & Mini-circuits & ZASWA-2-50DRA+ & DC to 5 GHz frequency range, absorptive SPDT Solid State Switch & 164 & 1 \\
		Circulator & Ditom & D3C2040 & 2 to 4 GHz frequency range & 450 & 1 \\
 \multicolumn{5}{l}{\textbf{Cables, connectors, terminators:}} \\
		N-type to SMA adapter & Mini-circuits & NM-SF50+ & Adapter for most signal generator's N-type output to SMA & 29 & 1 \\
		Terminator & Mini-circuits & ANNE-50+ & Terminator with SMA connection, for the unused microwave switch output & 14 & 1 \\
		High power terminator & MECA & 407-7 & High power terminator with SMA connection, for terminaton after the amplifier & 138 & 2 \\
	\end{tabular}
	\end{ruledtabular}
	\label{tab:microwave-equipment}
\end{table*}

\begin{table*}[h!]
	\centering
    \caption{(Table~\ref{tab:microwave-equipment} continued)}
	\begin{ruledtabular}
	\begin{tabular}{p{1.2in}p{0.8in}p{0.8in}p{2in}p{0.5in}p{0.2in}}
	Item & Vendor & Model & Description & Price (USD) & Qty \\
	\hline
		SMP/SMA cable & Mini-circuits & 047-12SMPSM+ & Flexible SMP/SMA cable to connect to and from custom PCB board & 36 & 2 \\
		MW cable, 3 ft & Mini-circuits & CBL-3FT-SMSM+ & Precision test cable with SMA connectors & 105 & 1 \\
		MW cable, 4 ft & Mini-circuits & CBL-4FT-SMSM+ & Precision test cable with SMA connectors & 114 & 1 \\
		MW cable, 6 ft & Mini-circuits & CBL-6FT-SMSM+ & Precision test cable with SMA connectors & 134 & 1 \\
		SMA-F/SMA-F adapter & Mini-circuits & SF-SF50+ & SMA female to SMA female adapter & 6 & 5 \\
		SMA-M/SMA-M adapter & Mini-circuits & SM-SM50+ & SMA male to SMA male adapter & 8 & 2 \\
		SMA-F/SMA-M adapter & Mini-circuits & SFR-SM50+ & Right angle SMA female to SMA male adapter & 35 & 2 \\
        PC/SMA Torque Wrench & Mouser & 74\_Z-0-0-21 & Huber + Suhner torque wrench for SMA connections & 157 & 1 \\
	\end{tabular}
	\end{ruledtabular}
	\label{tab:microwave-equipment_continued}
\end{table*}
\clearpage

\begin{table*}[h!]
	\centering
    \caption{Suggested components for single photon detection. Note that two APDs are required only for $g^{(2)}$ correlation measurements, and not used for any qubit manipulation experiments.}
	\begin{ruledtabular}
	\begin{tabular}{p{1.2in}p{0.8in}p{0.8in}p{2in}p{0.5in}p{0.2in}}
	Item & Vendor & Model & Description & Price (USD) & Qty \\
	\hline
		APD & Excelitas & SPCM-AQRH-43-FC & 250 cps dark counts,  22 ns dead time, 350 ps timing resolution,  400 to 1060 nm wavelength range, FC fiber adapter. 
        & 5816 & 1 or 2 \\
 \multicolumn{5}{l}{\textbf{Necessary components for $g^{(2)}$ measurement:}} \\
		Fiber splitter & Thorlabs & TW670R5F1 & 1x2 Wideband Fiber Optic Coupler, 670 $\pm$75 nm wavelength range, 50:50 Split, FC/PC & 388 & 1 \\
	\end{tabular}
	\end{ruledtabular}
	\label{tab:APD}
\end{table*}
\clearpage

\begin{table*}[h!]
	\centering
    \caption{Suggested components for permanent magnet and rotational control.}
	\begin{ruledtabular}
	\begin{tabular}{p{1.2in}p{0.8in}p{0.8in}p{2in}p{0.5in}p{0.2in}}
	Item & Vendor & Model & Description & Price (USD) & Qty \\
	\hline	
\multicolumn{5}{l}{\textbf{Mounts with 3 manual rotational axes:}} \\
            Micrometer Head & Starrett & T63XRL & 0-2" range, .0001" graduation & 329 & 1 \\
            Raw material for adapters & McMaster & 8975K322 & Multipurpose 6061 Aluminum 1" Thick X 4" Wide, 1 ft. Long & 50 & 2 \\
            Goniometer & Edmund & \#66-533 & 70mm, 150mm Radius, English Goniometer & 699 & 1 \\
            Goniometer & Edmund & \#66-534 & 70mm, 180mm Radius, English Goniometer & 727 & 1 \\
            Right-Angle Mounting Plate & Thorlabs & AP90 & 1/4"-20 Compatible & 95 & 1 \\
\multicolumn{5}{l}{\textbf{Magnet:}} \\
            Magnets & K\&J magnetics & DC8-N52 & 3/4" dia. x 1/2" thick, NdFeB grade N52 & 8 & 4 \\
	\end{tabular}
	\end{ruledtabular}
	\label{tab:magnet}
\end{table*}
\clearpage

\begin{table*}[h!]
	\centering
    \caption{Additional suggested components for NV center experiments.}
	\begin{ruledtabular}
	\begin{tabular}{p{1.2in}p{0.8in}p{0.8in}p{2in}p{0.5in}p{0.2in}}
	Item & Vendor & Model & Description & Price (USD) & Qty \\
	\hline
		High NA air objective & Olympus & MPLFLN100x & M Plan SemiApochromat, 100x/0.9 NA, 1 mm working distance & 2695 & 1 \\
		Objective for fiber collimation & Olympus & PLN10x & Plan Achromat Objective, 10x/0.25 NA, 10.6 mm working distance & 335 & 2 \\
		Thread adapter & Thorlabs & SM1A3 & Adapter with external SM1 threads and internal RMS threads, for Olympus objectives & 20 & 3 \\
		Dichroic beamsplitter & Semrock & Di03-R594-t3-25x36 & Dichroic used to overlap excitation and collection path. 594 nm cut-on wavelength & 660 & 1 \\
  	Rectangular Filter Mount & Thorlabs & FFM1 & 30-mm-cage-compatible rectangular filter mount for dichroic mirror & 68 & 1 \\
		Filter mount cage cube & Thorlabs & C4W & 30 mm cage cube to hold dichroic mirror & 70 & 1 \\
		Filter mount platform & Thorlabs & B3C & Fixed cage cube platform for C4W filter cage cube & 29 & 1 \\
		Long pass filter & Semrock & BLP01-633R-25 & Filters out excitation laser in collection path. 632.8 nm cut-on wavelength & 405 & 1 \\		
		Notch filter & Thorlabs & NF533-17 & Notch filter to filter out excitation light. Centered at 533 nm, FWHM = 17 nm (specific to excitation laser) & 594 & 1 \\
        
        \end{tabular}
	\end{ruledtabular}
	\label{tab:confocal}
\end{table*}

\begin{table*}[h!]
	\centering
    \caption{(Table~\ref{tab:confocal} continued)}
	\begin{ruledtabular}
	\begin{tabular}{p{1.2in}p{0.8in}p{0.8in}p{2in}p{0.5in}p{0.2in}}
	Item & Vendor & Model & Description & Price (USD) & Qty \\
	\hline
        Filter cage mount & Thorlabs & DCP1 & Drop-In 30 mm Cage Mount, Flexure Lock & 26 & 2 \\
		Broadband Dielectric Mirror & Thorlabs & BB1-E02 & 400 - 750 nm reflection range (two mirrors used in alignment procedure) & 81 & 6 \\
		Right-Angle Kinematic Mirror Mount & Thorlabs & KCB1C & Mirror mounts for 30 mm cage system. Note: mount has smooth cage rod bores & 155 & 4 \\
		Single Mode Patch Cable & Thorlabs & P1-630A-FC-2 & Fiber to direct collected NV center fluorescence to APD & 87 & 1 \\
		FC/PC Fiber Adapter Plate & Thorlabs & SM1FC & FC/PC Fiber Adapter Plate with External SM1 threads & 34 & 2 \\
		30 mm Cage Plate & Thorlabs & CP33 & Cage plate for optics, used with RMS thread adapter for excitation objective & 19 & 1 \\	
        XY Translating Lens Mount & Thorlabs & CXY1A & Adjust XY position of fiber coupling system in 30 mm Cage System & 201 & 2 \\
        Z-Axis Translation Mount & Thorlabs & SM1ZA & Adjust focus for fiber coupling system in 30 mm cage system & 220 & 2 \\
           Cap screw & Thorlabs & SH25S025 & 1/4"-20 Stainless steel cap screw, 1/4" long, 25 pack & 15 & 1 \\
           Cap screw & Thorlabs& SH25S050 & 1/4"-20 stainless steel cap screw, 1/2" long, 25 pack & 9 & 1 \\
            Washer & Thorlabs & W25S050 & 1/4" washer, M6 compatible, stainless steel, 100 Pack & 5 & 1 \\
	\end{tabular}
	\end{ruledtabular}
	\label{tab:confocal_continued}
\end{table*}

\begin{table*}[h!]
	\centering
    \caption{Suggested components for additional detectors in the excitation path.}
	\begin{ruledtabular}
	\begin{tabular}{p{1.2in}p{0.8in}p{0.8in}p{2in}p{0.5in}p{0.2in}}
	Item & Vendor & Model & Description & Price (USD) & Qty \\
	\hline
		Photo detector & Thorlabs & PDA8A2 & Detector that can be used to monitor excitation laser. & 485 & 1 \\
		Post & Thorlabs & TR1 & Ø1/2" optical post, L = 1", used for photodiode  & 5 & 1 \\
		Post holder & Thorlabs & PH1 & Ø1/2" optical post holder, L = 1", used for photodiode post & 8 & 1 \\
		Studded Pedestal Base Adapter & Thorlabs & BE1 & Ø1.25" studded pedestal base adapter, used for photodiode post  & 11 & 1 \\
		Clamping Fork & Thorlabs & CF175 & Clamping fork for Ø1/2", used for photodiode post & 12 & 1 \\
		CMOS camera & Thorlabs & CS165MU & Camera that can be used to view back-reflected image from the sample & 480 & 1 \\
		Camera lens & Thorlabs & MVL50M23 & 50 mm focal length, for 2/3" C-Mount Format Cameras & 233 & 1 \\
		Camera lens adapter & Thorlabs & SM1A10Z & Adapter with External SM1 Threads and Internal C-Mount Threads & 26 & 1 \\
		Beamsplitter cube & Thorlabs & BS010 & 50:50 Non-Polarizing Beamsplitter Cube used in excitation beam path to pick off laser beam and back-reflected image. & 179 & 1 \\
		Beamsplitter cube mount & Thorlabs & ARV1F & Rotatable Cage Mount & 118 & 1 \\
	\end{tabular}
	\end{ruledtabular}
	\label{tab:detectors}
\end{table*}

\begin{table*}[h!]
	\centering
    \caption{Suggested parts for breadboard and vibration isolation table.}
	\begin{ruledtabular}
	\begin{tabular}{p{1.2in}p{0.8in}p{0.8in}p{2in}p{0.5in}p{0.2in}}
	Item & Vendor & Model & Description & Price (USD) & Qty \\
	\hline
		Breadboard & Thorlabs & B3036G & 30" x 36" honeycomb breadboard with 1/4-20" threaded holes used to fasten optics posts. & 1627 & 1 \\
		Vibration Isolation Table & Thorlabs & PFA51505 & Active isolation frame to use with 30" x 36" breadboard. & 3035 & 1 \\
		Casters for Table Frame & Thorlabs & PWA061 & 4" Caster Kit, Set of 4 & 144 & 1 \\
		Air Filter/Regulator & Thorlabs & PTA013 & Air Filter / Regulator & 89 & 1 \\
	\end{tabular}
	\end{ruledtabular}
	\label{tab:table}
\end{table*}

\begin{table*}[h!]
	\centering
    \caption{Suggested components for 30 mm cage system.}
	\begin{ruledtabular}
	\begin{tabular}{p{1.2in}p{0.8in}p{0.8in}p{2in}p{0.5in}p{0.2in}}
	Item & Vendor & Model & Description & Price (USD) & Qty \\
	\hline
		Mounting Post & Thorlabs & P1.5 & Ø1.5" Mounting Post, 1/4"-20 Taps, L = 1.5". Supporting the four kinematic mirror mounts and camera. & 27 & 5 \\
		Studded Pedestal Base Adapter & Thorlabs & PB4 & Ø1.85" Studded Pedestal Base Adapter, 1/4"-20 Thread & 15 & 5 \\
		Clamping Fork & Thorlabs & PF125B-P5 & Clamping Fork for Ø1.5" Pedestal Post or Post Pedestal Base Adapter, 5 Pack & 75 & 1 \\
		Cage Rod & Thorlabs & ER1-P4 & Cage Assembly Rod, 1" Long, Ø6 mm, 4 Pack & 21 & 4 \\
		Cage Rod & Thorlabs & ER3-P4 & Cage Assembly Rod, 3" Long, Ø6 mm, 4 Pack & 27 & 2 \\
		Cage Rod & Thorlabs & ER4-P4 & Cage Assembly Rod, 4" Long, Ø6 mm, 4 Pack & 29 & 1 \\
       \end{tabular}
    \end{ruledtabular}
	\label{tab:cage_system}
\end{table*}
  
\begin{table*}[h!]
	\centering
    \caption{Suggested components for optical enclosure. The hardboard and rail extrusions will need to be cut to your chosen dimensions. The dimensions of the optical enclosure in Fig.~\ref{fig:confocal} are 36 in. x 30 in. x 6 in. Additionally, the dimensions of the APD enclosure are 9 in. x 6 in. x 4.5 in}
	\begin{ruledtabular}
	\begin{tabular}{p{1.2in}p{0.8in}p{0.8in}p{2in}p{0.5in}p{0.2in}}
	Item & Vendor & Model & Description & Price (USD) & Qty \\
	\hline
		Black Hardboard & Thorlabs & TB4 & Light-blocking material for optical enclosure sides & 76 & 2 \\
		Black masking tape & Thorlabs & T137-1.0 & Recommended to tape corners and other joints to eliminate the passage of light through those areas and reduce shedding from inner foam, 1" x 180' & 11 & 1 \\
		Rail Extrusion & Thorlabs & XE25RL2 & Rails for frame of optical enclosure & 46 & 2 \\
		Right angle brackets & Thorlabs & XE25A90 & Fasten rails of optical enclosure to optical table & 32 & 4 \\
		Construction Cube & Thorlabs & RM1S & Construction cubes for corners of optical enclosure lid & 23 & 8 \\
		Low-Profile Channel Screw & Thorlabs & SH25LP63 & Low-profile screws for optical enclosure construction & 22 & 1 \\
		Blackout fabric & Thorlabs & BK5 & Black Nylon, Polyurethane-Coated Fabric, 5' x 9', useful to cover fiber to APD & 63 & 1 \\
       \end{tabular}
    \end{ruledtabular}
\label{tab:optical_enclosure}
\end{table*}

\begin{table*}[h!]
	\centering
    \caption{Suggested tools for optics lab and alignment procedures}
	\begin{ruledtabular}
	\begin{tabular}{p{1.2in}p{0.8in}p{0.8in}p{2in}p{0.5in}p{0.2in}}
	Item & Vendor & Model & Description & Price (USD) & Qty \\
	\hline	
		Cage Alignment Plate & Thorlabs & CPA1 & 30 mm Cage Alignment Plate with Ø0.9 mm Hole & 15 & 3 \\
		Pivoting Drop-In Optic Mount & Thorlabs & 	
        CP360Q & 30 mm cage optics mount, useful to direct beam out of cage system for alignment & 111 & 1 \\
        Balldrivers and hex keys & Thorlabs & TC4 & 18-piece, breadboard mountable ball driver and tool caddy kit, imperial & 86 & 1 \\
        Spanner Wrench & Thorlabs & SPW602 & Spanner wrench for SM1-threaded retaining Rings & 31 & 1 \\
        Magnetic Beam Height Ruler & Thorlabs & BHM3 & 6" magnetic beam height measurement tool & 32 & 2 \\
        Lens Tissues & Thorlabs & MC-5 & Lens tissues, 25 sheets per booklet, 5 booklets & 12 & 1 \\
        Fiber Connector Cleaner & Thorlabs & FCC-7020 & 20' spool & 27 & 1 \\
        Slotted Tube & Thorlabs & SM1L20C & Slotted lens tube used in alignment as a stand-in objective & 73 & 1 \\
        Irises & Thorlabs & SM1D12 & Irises to attach to slotted lens tube, used for alignment procedure & 68 & 1 \\
        Visual Fault Finder & Jonard Tools & VFL-25 & 650 nm wavelength fault finder with FC/PC output, used as red laser source for alignment procedure & 128 & 1 \\
        Laser goggles & Thorlabs & LG3 & 180 to 532 nm, OD = 7+ & 178 & 4 \\
	\end{tabular}
	\end{ruledtabular}
\label{tab:lab tools}
\end{table*}
\clearpage

\section{Design files and example code\label{appendix:example_code}}
Design files for the setup and example instruction code can be found on GitHub:\\ \url{https://github.com/nvsareforever/NV-lab-example-code}

\end{document}